\documentclass[aps,preprint,showpacs,superscriptaddress,groupedaddress]{revtex4}  
\usepackage{graphicx}  
\usepackage{dcolumn}   
\usepackage{bm}        
\usepackage{amssymb}   
\usepackage{color}
\usepackage{hyperref}

\begin{document}

\widetext


\title{Structural characterizations of water-metal interfaces}
\author{Kevin~Ryczko} \affiliation{Department of Physics, University of Ontario Institute of Technology, Oshawa, Ontario, L1H 7K4}
\author{Isaac~Tamblyn} \affiliation{Department of Physics, University of Ontario Institute of Technology, Oshawa, Ontario, L1H 7K4} 
\affiliation{Department of Physics, University of Ottawa, Ottawa, Ontario, K1N 6N5}
\affiliation{National Research Council of Canada, 100 Sussex Drive, Ottawa, Ontario, K1A 0R6}                            
\date{\today}

\begin{abstract}
We analyze and compare the structural, dynamical, and electronic properties of liquid water next to prototypical metals including Pt, graphite, and graphene. Our results are built on Born-Oppenheimer molecular dynamics (BOMD) generated using density functional theory (DFT) which explicitly include van der Waals (vdW) interactions within a first principles approach. All calculations reported use large simulation cells, allowing for an accurate treatment of the water-electrode interfaces. We have included vdW interactions through the use of the optB86b-vdW exchange correlation functional. Comparisons with the Perdew-Burke-Ernzerhof (PBE) exchange correlation functional are also shown. We find an initial peak, due to chemisorption, in the density profile of the liquid water-Pt interface not seen in the liquid water-graphite interface, liquid water-graphene interface, nor interfaces studied previously. To further investigate this chemisorption peak, we also report differences in the electronic structure of single water molecules on both Pt and graphite surfaces. We find that a covalent bond forms between the single water molecule and the Pt surface, but not between the single water molecule and the graphite surface. We also discuss the effects that defects and dopants in the graphite and graphene surfaces have on the structure and dynamics of liquid water.

\end{abstract}

\maketitle

keywords: water-solid interface, fundamental surface chemistry, molecular dynamics, molecular structure, electronic structure.

\section{Introduction}
Examining the molecular structure of water-solid interfaces is crucial for understanding dynamical processes that occur in both natural and controlled environments. Hydrogen production through the splitting of water at a metal surface is of great interest for solar cell devices \cite{walter2010solar}, and generating hydrogen and oxygen gas has been proposed as a means to store energy \cite{lewis2006powering}. Despite significant efforts both computationally \cite{cicero2008water, carrasco2013role, gross2014water,li2012influence, lin2005simulation, liu2008density, liu2011initial,liu2010structure, ma2011adsorption, beret2008aqueous, michaelides2006density, chen2012hydrogen, zhao2007wetting, velasco2014structure} and experimentally \cite{jiang2014real, chen2012hydrogen, toney1994voltage, willard2009water} to study electrode-water interfaces, to-date, there has yet to be an explicit treatment of liquid water next to a realistic, catalytic surface computed at the level of accurate, first principles molecular dynamics. Almost 10 years ago, Cicero {\it{et al.}} \cite{cicero2008water} studied several liquid water-graphene interfaces as well as water confined in carbon nanotubes with differing radii using DFT with the PBE exchange correlation functional \cite{perdew1996generalized}. Despite the differences of the simulation cell sizes, the interfacial water had a similar structure when comparing the density profiles of the different supercells. Moving away from the graphene at a perpendicular direction, a zero particle density was found for $\approx2.5$ \AA. Further past this volume of zero particle density, away from the surface, an initial spike was seen where the particle density was significantly larger than the density of bulk water in ambient conditions. In a subsequent study by Liu {\it et al.} \cite{liu2011water}, a liquid water-salt interface was examined using the same theoretical framework. Here, the structure of the interfacial layer was quite different. Despite having a similar zero particle density adjacent to the surface, the first peak in the density was lower, and more broad. Even though graphite has an inferior efficiency for water splitting when compared to Pt, it is commonly used in laboratories to study electrochemical processes because it is inexpensive and abundant. By studying the structural and electronic properties of liquid water-solid interfaces through {\it ab initio} methods, we can begin to understand and characterize electrodes. This aids in the selection process of a new, abundant, and efficient alternative electrode to superior rare metals. 

More recently, another {\it ab initio} study was done by Velasco {\it et al.} \cite{velasco2014structure} on a liquid water-Au interface. Here, the computational efforts were exceptional; 80 ps of MD trajectories were generated using DFT with the PBE exchange correlation functional. The molecular density profile of this interface was similar to the liquid water-graphene interface; the only structural difference was an additional second peak $\approx5$ \AA~ past the initial, most prominent peak. It should also be noted that the density of the initial, largest peak reached a value of 4 g/cc; this is approximately twice the value of the largest peak seen for the water-graphene interface. Thus, the water shows more attraction to the gold surface than the graphene, but it may not be a key indicator of an efficient catalyst. Gold is known to be chemically inert \cite{hammer1995gold}. When a molecule (H${}_2$ for example) binds to the surface of gold, the energies associated with the anti-bonding molecular orbitals lie below the Fermi level. This results in occupancy of the anti-bonding molecular orbitals, causing repulsion and reducing surface activity.

The liquid water-solid interfaces considered in our work are Pt, graphite, and graphene. Pt is a rare metal, and is known to have a low over potential for water splitting, whereas carbon-based materials are less efficient. Due to the efficiency, rarity, and cost of Pt, there is an active area of research focused on finding an alternative, abundant metal/semi-metal with similar properties \cite{reece2011wireless, orlandi2010ruthenium, du2012elucidating,du2012catalysts,liu2011water}. Many theoretical and experimental studies have been done to further understand the water-Pt interface \cite{xia1995electric, willard2013characterizing, osawa2008structure, smith1988ultrathin, heinzinger1991molecular, ogasawara2002structure, limmer2015water, otani2008structure}. In particular, Osawa {\it et al.} \cite{osawa2008structure} performed infrared absorption spectroscopy experiments to examine the structure of water next to an electrified, pristine Pt electrode. They found significant differences in the spectrum when comparing the interfacial water to that of the bulk. They concluded that the differences in the spectrum were due to the orientation of the surface water molecules. At the positively charged electrode, the surface molecules formed a strong hydrogen-bonding network parallel to the surface. At the negatively charged electrode, the interfacial water formed a much weaker hydrogen bonding network and the water molecules preferred an H-down orientation to the surface. The majority of computational studies regarding the water-Pt interface have relied on the use of classical force fields (coupled with molecular dynamics or Monte Carlo engines). They have reported statistical distributions to understand the structure of water next to the surface \cite{xia1995electric, nagy1990molecular, heinzinger1991molecular}, such as the molecular density profile, orientation probability distributions, and two-dimensional probability distributions outlining adsorption sites. More recent studies have used experimental and \emph{ab initio} techniques to study this interface \cite{bergès2013quantum, sberegaeva2014mechanistic, yeh2013molecular}, but for smaller systems which only included a surface layer (excluding bulk water). In this report, we compare our results to previous water-solid studies and build on previous classical reports of the liquid water-Pt interface with a quantum description of the forces which define this interface. A larger system size allows for dynamics in the liquid that would occur naturally in ambient conditions; the natural dynamics we encapsulate in our simulations are absent in previous simulations that only consider the interfacial surface layer of water.

This report is split into four sections. In section II, we discuss our method for generating molecular dynamics trajectories. In section III, we assess the importance of exchange-correlation functional choice, comparing functionals with (optB86b-vdW \cite{klimevs2011van}) and without (PBE) vdW interactions. In section IV we compare catalytically active (Pt) and inert graphitic surfaces. Lastly, in section V, we discuss the effects of adding defects and dopants into graphitic surfaces.

\section{Methods}
\begin{figure}[h!]
\centering
\includegraphics[width=\linewidth]{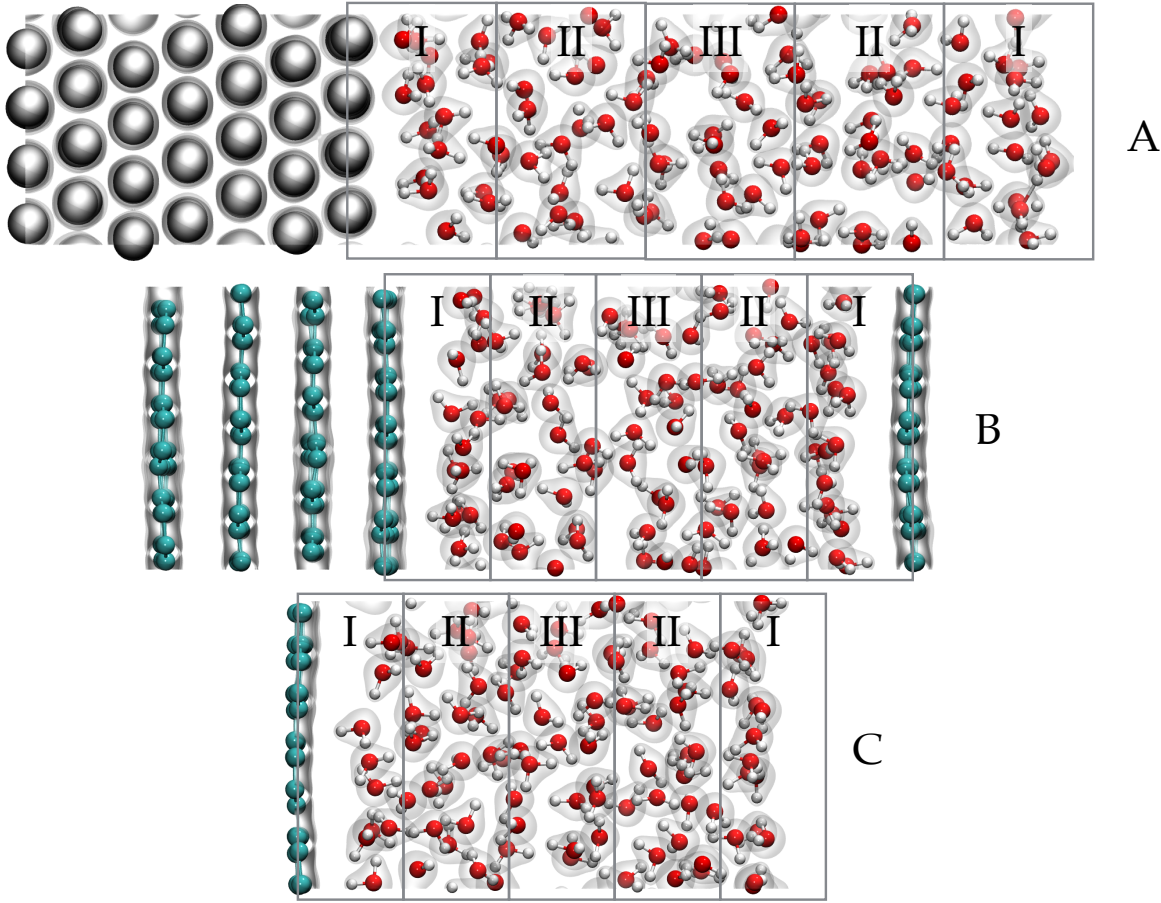}
\caption{\label{f:supercell} Visualizations of the liquid water-Pt interface (A), liquid water-graphite interface (B), and liquid water-graphene interface (C) used in the DFT calculations. Here, the transparent surfaces around the atoms are the computed self-consistent charge densities for the geometries seen.}
\end{figure}

\begin{figure}[h!]
\centering
\includegraphics[width=0.32\linewidth]{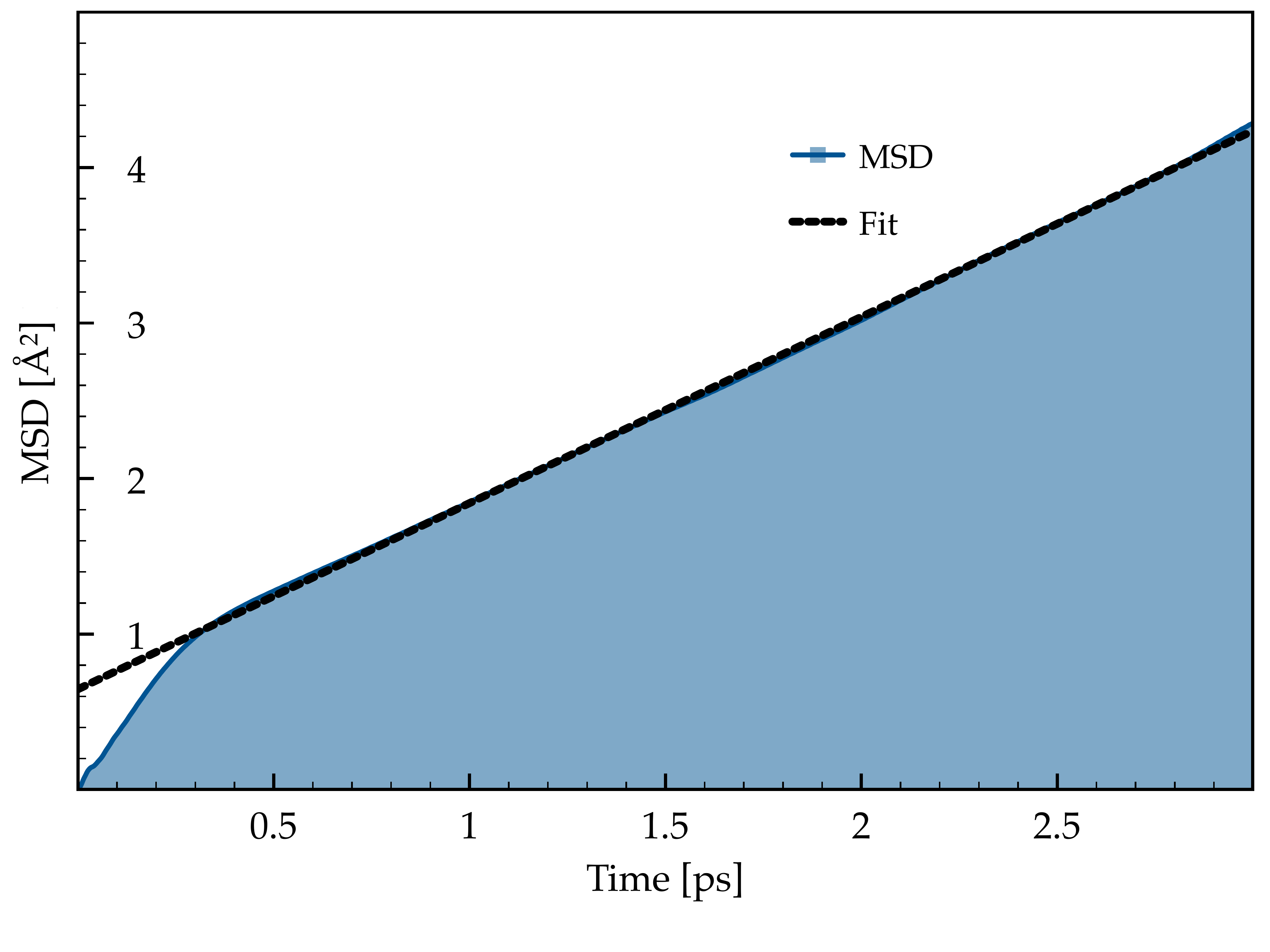}
\includegraphics[width=0.32\linewidth]{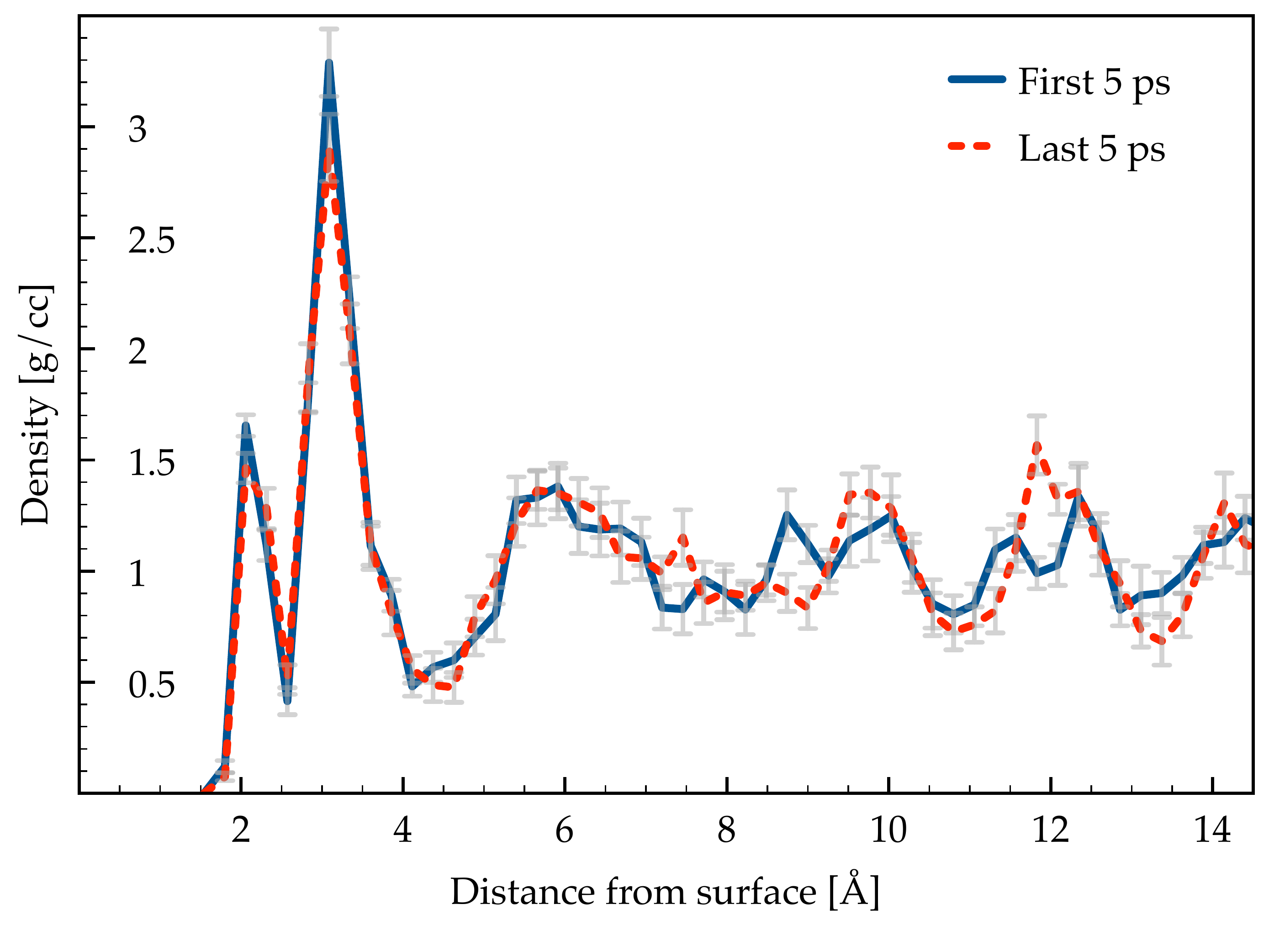}
\includegraphics[width=0.32\linewidth]{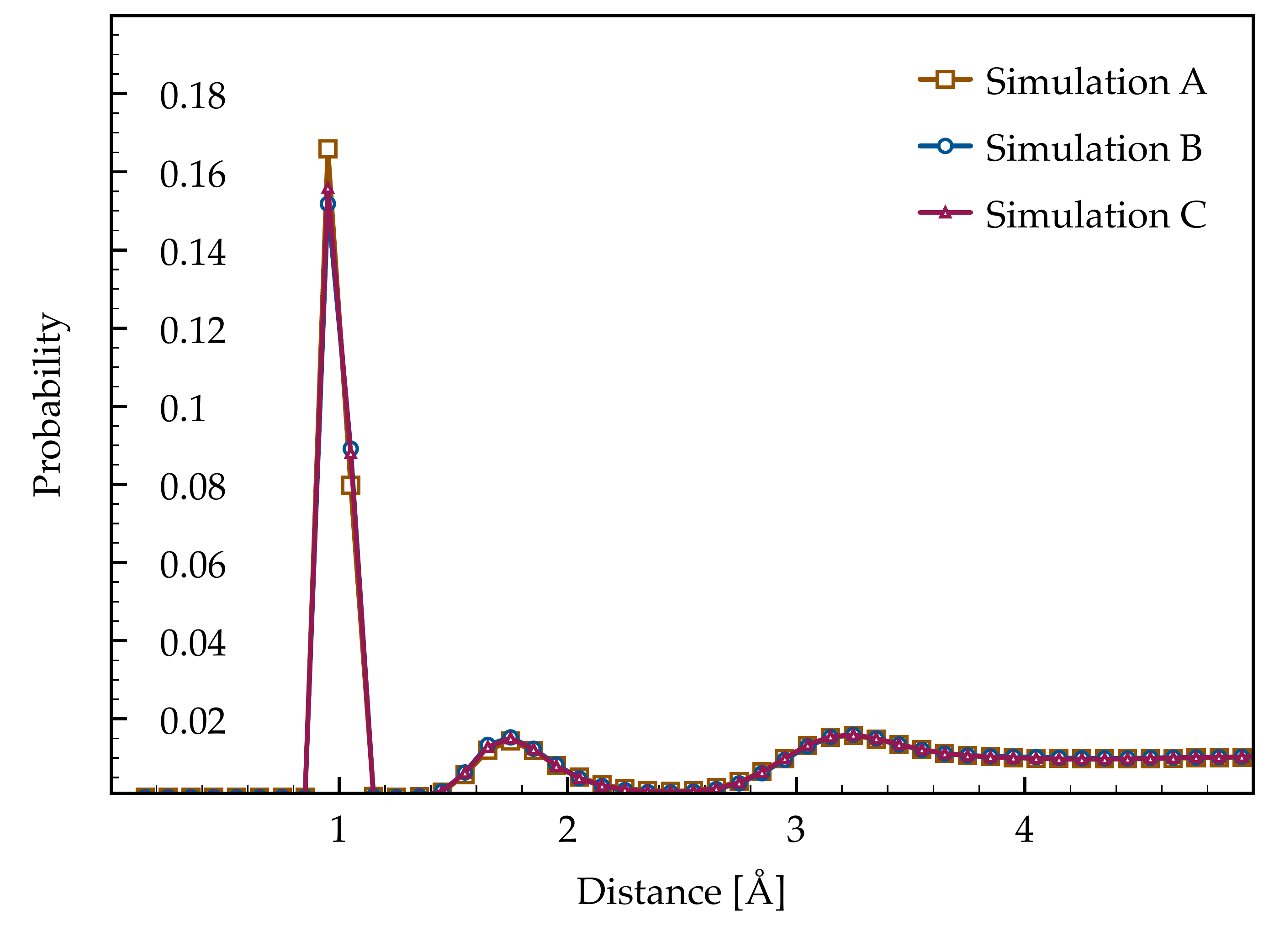}
\caption{\label{f:convergence_figure}Left: Mean squared displacement of 100 water molecules in a cubic supercell ($14.41^3$ \AA${}^3$) with a thermostat of 400 K using the vdW-optB86b exchange correlation functional. The fit of the asymptotic behaviour gives a diffusion coefficient of $2.0\times10^{-9}$ m${}^2$s${}^{-1}$, which is slightly lower than the experimental value of $2.3\times10^{-9}$ m${}^2$s${}^{-1}$. Middle: Molecular density profiles of a 10 ps long simulation (with a thermostat of 400 K) of a liquid water-Pt interface using the vdW-optB86b exchange correlation functional. 10 ps of MD was generated, and then divided in half to give two trajectories, each 5 ps long labelled as `First 5 ps' or `Last 5 ps'. The error bars are the standard error found from analyzing the molecular density profile at each time step. Right: O-H radial distribution functions for liquid water-Pt interfaces at 330 K using the PBE functional. 3 independent MD trajectories, each 5 ps long, were generated and are labelled as simulation A, B, and C. Here, it is clear that the O-H radial distribution functions have converged when comparing the different simulations.}
\end{figure}

Here, we provide the preparation steps followed to generate all of the water-solid interfaces used in our analysis. For the liquid water-graphite and liquid water-graphene interfaces, we used supercell dimensions $12.41 \times 12.81 \times 37.36$ \AA\hspace{1mm} and $12.41 \times 12.81 \times 23.96$ \AA\hspace{1mm} respectively. For the liquid water-graphite interface, we used 300 C atoms to form 5 layers. In the liquid water-graphene interface, we used 60 C atoms. In addition to the pristine surfaces, we have also constructed interfaces with defects or dopant materials. To model defects, we considered two classes. In the first class, we removed 2 carbon atoms from the surface layer(s) of both graphene and graphite. In the second class, we modified the crystalline structure at the surface(s) of graphene and graphite to create Stone-Wales (5-7) defects. To model doping, we replaced a surface carbon atom with either a nitrogen or boron atom in the surface layer(s). For all liquid water-graphitic interfaces, we used 100 D${}_2$O molecules, a $1\times1\times1$ k-point grid centered about the $\Gamma$ point, and a plane wave energy cutoff of 500 eV. We used the exchange-correlation functional optB86b-vdW and a Nos\'{e}-Hoover thermostat (T = 300, 330, and 400 K). For the liquid water-Pt interface, we constructed two distinct Pt(111)-D$_2$O interfaces, both with supercell dimensions of $11.24\times9.74\times45.69$ \AA, 112 Pt atoms, and 100 D$_2$O molecules. For one of the interfaces, we used the PBE exchange correlation functional, and for the other interface, we used the optB86b-vdW  exchange correlation functional. We found that both the PBE and optB86b-vdW exchange correlation functionals give the same lattice constant for bulk Pt. When using the optB86b-vdW functional, we ran molecular dynamics at T = 300, 330, 400, 500, 600, 650, 700, 750, 800, and 1000 K. When using the PBE exchange correlation functional, we ran molecular dynamics at T = 330, 400, and 1000 K. For both interfaces, we used a plane wave energy cutoff of 500 eV as well as a $2\times2\times1$ k-point grid centered about the $\Gamma$ point. Figure \ref{f:supercell} shows the supercells used for the simulations. When constructing the liquid water-Pt interfaces, we chose to use 7 layers of Pt with the 3 middle layers constrained along the $z$ axis. Since we used periodic boundary conditions, the liquid water interacts with two metal faces (assuming no vacuum layer). The two faces should be identical such that the supercell is symmetric. It should also be noted that when placing the water next to the metal surface, an exclusion volume must be included. This is a volume of empty space that occurs naturally and is dependent on the metal. When Cicero \emph{et al.} \cite{cicero2008water} constructed their liquid water-graphene interface, the thickness (in the $z$ direction) of the exclusion volume was $\sim$ 2 \AA. For the liquid water-Pt interfaces, we chose the thickness of the exclusion volume to be $\sim$ 2.35 \AA. For the liquid water-graphite and liquid water-graphene interfaces, we chose the thickness of the exclusion volume to be $\sim$2.5 \AA. After the supercells were constructed, we then structurally relaxed the systems to eliminate the random configurations of the water molecules. Afterwards, we ran molecular dynamics and monitored the pressure, temperature, and total energy of the systems. Even after the structural relaxations, a few ps of data showed significant fluctuation as the systems approached their equilibrium states. This data was not considered in our analysis. Once the systems reached their equilibrium states, we collected 10 ps of data for analysis.  We found that the structure of the density profile converges at $\approx$ 10 ps, which can be seen in Figure \ref{f:convergence_figure}. To about 5 \AA\hspace{1pt} away from the surface, the water has a similar structure. Beyond 5 \AA, the water fluctuates about 1.1 g/cc. The same amount of fluctuation can be seen when comparing different MD trajectories of pure water. The convergence of the density profile is much slower than other distributions. In Figure \ref{f:convergence_figure}, we also show the O-H radial distribution function which is well converged through 5 ps. For some of our structural analysis, we have also included error bars which give the standard error of the mean. Subsequent snapshots of molecular dynamics are highly correlated. To remove the correlation, we performed our analysis such that statistical distributions were recursively split and averaged over (block averaging). All of the molecular dynamics reported were done using the canonical (NVT) ensemble, heavy water (D${}_2$O), and Nos\'{e}-Hoover thermostats. The use of heavy water allowed for the increase of the time step by 40\% (8 a.u. to 11.3 a.u.). Some of the NVT calculations were also replicated in the microcanonical (NVE) ensemble to verify that the thermostat did not bias the results. We used the Vienna {\it ab initio} Simulation Package (VASP) \cite{kresse1996software} for all simulations. In total, over 250 ps of BOMD was generated for this report.

Additional total energy calculations were also run for the water-graphite and water-Pt interfaces where only one water molecule was considered on the surface. The supercells had the same cell dimensions as mentioned above. For these calculations, structural minimizations were done using a $4\times 4\times 2$ k-point grid centered about the $\Gamma$ point and an atomic force convergence of $10^{-4}$ eV/\AA. The exchange-correlation functional used was optB86b-vdW.

In our analysis, we chose a minimum temperature of 400 K for the molecular dynamics due to earlier studies done on pure liquid water \cite{schwegler2004towards, pham2016structure}. Schwegler \emph{et al.} \cite{schwegler2004towards} found when using the PBE exchange correlation functional at 300 K, the water is overstructured after examining radial distribution functions, and the diffusion coefficient is lower than the reported value using experimental techniques. They found that increasing the temperature approximated the inclusion of proton quantum effects in their calculations. In a study done by Morales \emph{et al.} \cite{morales2013towards}, they found that excluding the nuclear quantum effects of ions leads to artificially low displacements at low temperatures. In Figure \ref{f:convergence_figure}, we show the mean squared displacement for liquid water in a cubic cell at 400 K. Here, the diffusion coefficient was calculated to be $2.0\times10^{-9}$ m${}^2$s${}^{-1}$, which is slightly lower than the experimental value of $2.3\times10^{-9}$ m${}^2$s${}^{-1}$ found from neutron scattering \cite{spyroudiffusion}.
\begin{figure}
\centering
\includegraphics[width=\linewidth]{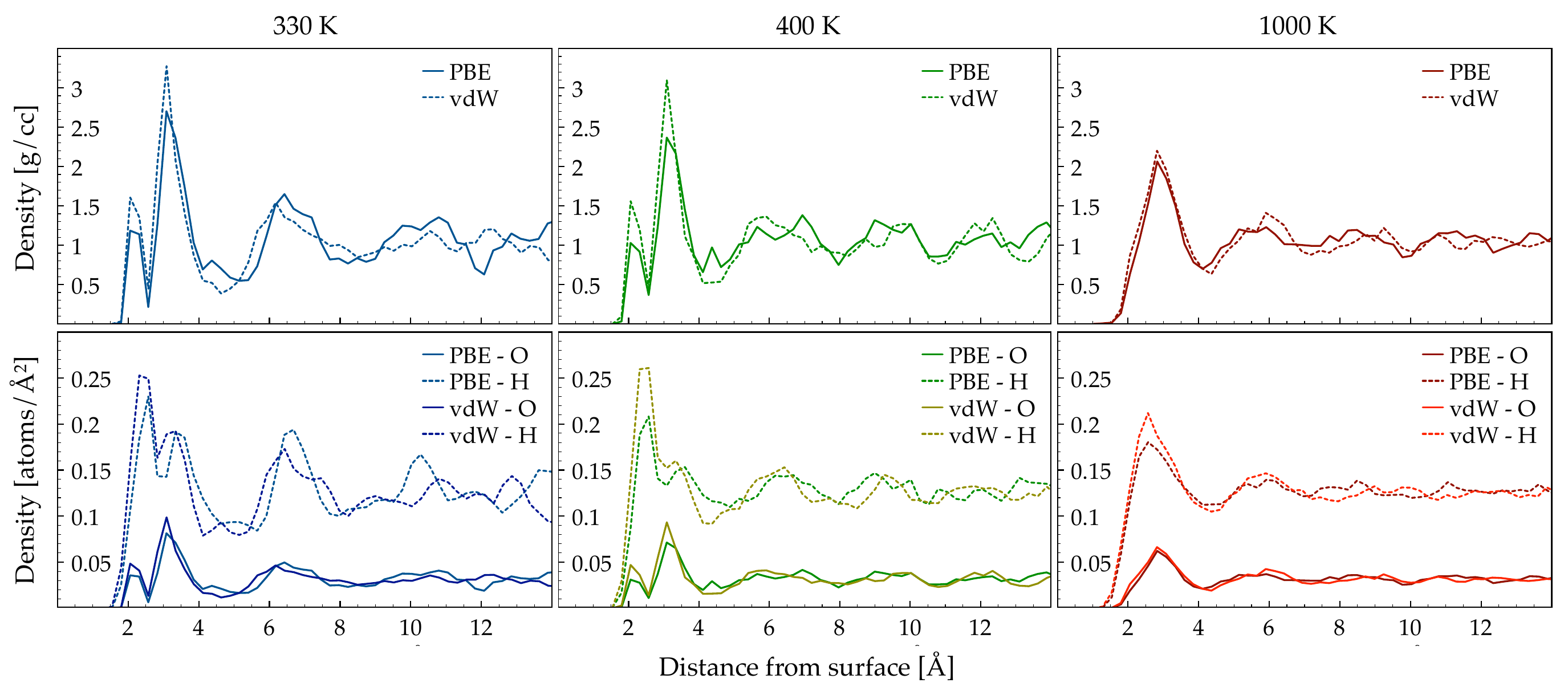}
\caption{\label{f:main_dens_profile} Molecular (top 3 figures) and atomic (bottom 3 figures) density profiles for the liquid water-Pt interfaces at 330 K, 400 K, and 1000 K. The top two curves in the molecular density profiles are from two independent MD trajectories, one with vdW interactions (labelled as vdW) and one without (labelled as PBE). The bottom 4 curves are from the same MD trajectories, but have been split up by either O or H atoms to give atomic density profiles.}
\end{figure} 

\section{Assessing the importance of van der Waals interactions}
Recently, it has been shown that vdW forces have a significant contribution in the interaction between water molecules and metals. Carrasco {\it et al.} \cite{carrasco2013role} calculated the adsorption energy of a water molecule on a Pt surface and found that it is almost halved when excluding such interactions (-403 meV with vdW interactions (optB88-vdW) and -217 meV without (PBE)). Reports with only one or two layers of water molecules next to a metal surface (ice-metal interfaces) are common in the literature \cite{ludwig2003does, sexton1980vibrational, michaelides2001catalytic} due to their smaller system sizes and lower computational cost. A computational study of the magnitude we present is missing. We begin our study by examining the structure of the water molecules next to the Pt surface. In Figure \ref{f:main_dens_profile}, we plot the molecular and atomic density profiles as a function of distance from the nearest Pt surface. Immediately adjacent to the surface, we find high density water. This is a common feature of water-electrode interfaces, and has been found using X-ray scattering \cite{toney1994voltage} and {\it ab initio} methods \cite{cicero2008water, velasco2014structure, liu2011water}. The first notable result is the difference between the profiles when comparing the PBE functional with the optB86b-vdW functional. For the optB86b-vdW functional, the largest peaks have heights 15-29\% greater than the largest peaks for the PBE functional (depending on temperature). This suggests that there is an increased attraction between the Pt atoms and water molecules when vdW interactions are introduced. This is consistent with the work done by Carraso {\it et al.} \cite{carrasco2013role}. The second notable result is the appearance of an initial wetting peak before the largest peak in the density. This peak has not been seen in density profiles of other water-solid interfaces \cite{cicero2008water, velasco2014structure, liu2011water}, and is absent in the liquid water-graphite and liquid water-graphene interfaces covered in this report. This initial peak is from water molecules chemisorbing to the surface. Past experimental work done using electron energy loss spectroscopy \cite{sexton1980vibrational}, and ultraviolet/X-ray photoemission \cite{fisher1980interaction} also found chemisorption at low temperatures. When increasing the temperature of the system, the peak remains at 400 K, but vanishes at 1000 K. This suggests that the energy of the Pt-H${}_2$O bond is somewhere in the range of $k_BT$ where $400 \text{ K}<T<1000 \text{ K}$. We ran subsequent molecular dynamic simulations with a thermostat of 500, 600, 650, 700, 750, and 800 K; we find that the initial wetting peak starts to decrease in size around 600 K, and is almost completely gone at 800 K. This is consistent with previous thermal desorption spectroscopy experiments done for oxygen-Pt \cite{gland1978adsorption} and hydrogen-Pt \cite{christmann1976adsorption} interfaces. Here, it was found that hydrogen desorbs around 420 K, but oxygen desorbs at 850 K. At higher temperature (i.e. beyond the boiling point), the system is clearly in a non-equilibrium state, but still gives insight into the nature of the chemisorption layer.


\begin{figure}[h!]
\centering
\includegraphics[width=\linewidth]{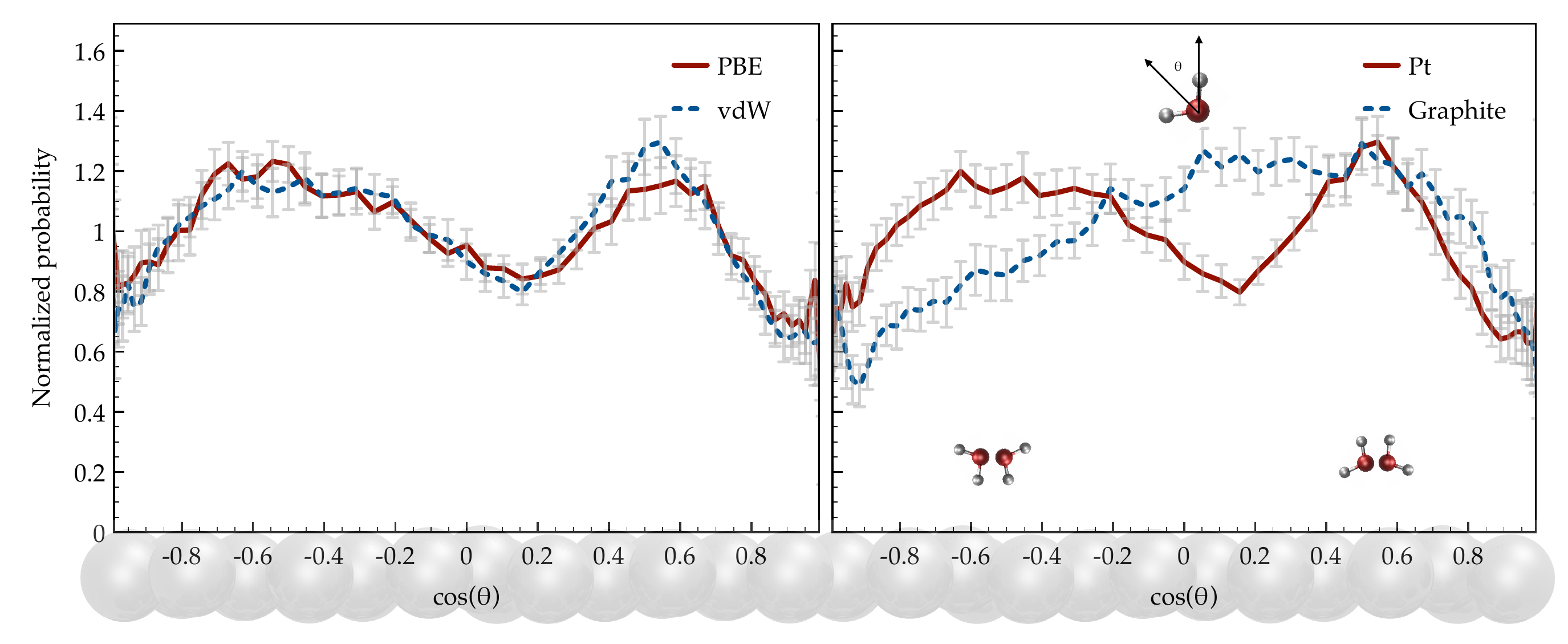}
\caption{\label{f:pt_orient} Normalized probability distributions of a water molecule having a certain orientation. In Figure \ref{f:supercell}, we divided the supercell into regions (I, II, and III) separating bulk from surface water molecules. Here, we only show region I; region II and III show little structure and a more random orientation. Here, $\theta$ is the angle between the geometric dipole to the $z$-normal. In the left figure, we compare 2 independent water-Pt interfaces; one with vdW interactions (labelled as vdW) and one without (labelled as PBE). In the right figure, we compare a water-Pt (labelled as Pt) and water-graphite interface (labelled as Graphite), both with vdW interactions. }
\end{figure}

In Figure \ref{f:pt_orient}, we look at the normalized probability of finding a surface water molecule with a certain orientation. To normalize, we first considered the orientation probability distribution of a randomly oriented water molecule. We then calculated the orientation probability distribution for a particular system and divided by the randomly oriented probability distribution. This is analogous to the normalization procedure in the calculation of a radial distribution function. In Figure \ref{f:supercell}, we outline regions within the supercell to differentiate surface water molecules from bulk ones. For both the PBE and optB86b-vdW simulations, there is a similar trend for surface water molecules. The water molecules orient themselves so that either an oxygen (O-Pt bond) or hydrogen (H-Pt bond) faces the surface. In another study done by Carrasco \emph{et al.} \cite{carrasco2009insight}, they found that the preferred orientation for a water molecule sitting on 4d metal surfaces (Ru(0001), Rh(111), Pd(111), Ag(111)) is with the geometric dipole almost flat to the surface and the oxygen over the atop site. When bulk water is introduced, surface water molecules bond with other water molecules within the surface layer, or other water molecules further away from the surface (in the bulk). Due to the interaction with the metal surface as well as bulk water, the surface water molecules constantly reorientate themselves to bind with the surface as well as adjacent water molecules. The environment in the interfacial layer is highly competitive. Noting the differences between the PBE and the optB86b-vdW simulations, for PBE, there is a much sharper peak where the oxygen atom is facing towards the surface. In Table \ref{t:OdownHdown}, we highlight the probability of finding water molecules (within a distance of 2.5 \AA~and 5.0 \AA~from the surface) with an O-down (both hydrogens further away from the surface), O-up (both hydrogens closer to the surface), or mixed orientation. Interestingly, there is very little difference when comparing the functional choice. Closer to the surface ($<$ 2.5 \AA~away), no water molecule has an O-up orientation. The water molecules will either have an O-down, or mixed orientation with equal probability. These results are similar to the results found by Velasco {\it et al.} \cite{velasco2014structure} for a liquid water-Au interface. They found that 49\% of the time surface water molecules orientate themselves to have an orientation where the geometric dipole of the water molecule is parallel with the surface. In this orientation, the lone-pairs of the oxygen interact with the surface orbitals, and the hydrogens form intermolecular bonds with adjacent molecules. This could be thought of in our framework as the O-down orientation. Velasco \emph{et al.} also found that 49\% of the time, water molecules would orientate themselves to have one hydrogen facing the surface, and the other participating in the hydrogen bonding network. This is analogous to our mixed orientation. Along with the theoretical study done by Velasco \emph{et al.}, X-ray absorption spectroscopy revealed that unsaturated hydrogen bonds occur at the surface. The changes in the spectrum are simply due to the reorganization of water molecules at the surface once the electrode is introduced. Looking further away from the surface ($<$ 5.0 \AA~away), the O-down and mixed probabilities decrease equally, and O-up molecules are seen with a low frequency.\\
\renewcommand{\thefootnote}{\roman{footnote}}
\begin{center}
\begin{table}[h]
\begin{tabular}{c|c|c|c}
	Description & O-down probability $\pm$ $\sigma$ & O-up probability $\pm$ $\sigma$ & Mixed probability $\pm$ $\sigma$ \\\hline
	Pt (PBE - 2.5 \AA) & 0.50 $\pm$ 1.6 $\times10^{-5}$ & 0.0 & 0.50 $\pm$ 1.6 $\times10^{-5}$ \\
    Pt (PBE - 5.0 \AA) & 0.45 $\pm$ 2.2 $\times10^{-3}$ & 0.10 $\pm$ 4.3 $\times10^{-3}$ & 0.45 $\pm$ 2.2 $\times10^{-3}$ \\    
	Pt (vdW - 2.5 \AA) & 0.50 $\pm$ 7.2 $\times10^{-5}$ & 0.0 & 0.50 $\pm$ 7.6 $\times10^{-5}$ \\  
    Pt (vdW - 5.0 \AA) & 0.45 $\pm$ 2.2 $\times10^{-3}$ & 0.10 $\pm$ 4.5 $\times10^{-3}$ & 0.45 $\pm$ 2.2 $\times10^{-3}$ \\
	Graphite (2.5 \AA) & 0.50 $\pm$ 6.0 $\times10^{-3}$ & 0.0 & 0.50 $\pm$ 6.1 $\times10^{-3}$ \\
	Graphite (5.0 \AA) & 0.44 $\pm$ 1.9 $\times10^{-3}$ & 0.12 $\pm$ 3.8 $\times10^{-3}$ & 0.44 $\pm$ 1.9$\times10^{-3}$ \\
	Graphene (2.5 \AA) & 0.50 $\pm$ 4.9 $\times10^{-3}$ & 0.0 & 0.50 $\pm$ 4.9 $\times10^{-3}$ \\
	Graphene (5.0 \AA) & 0.43 $\pm$ 1.8 $\times10^{-3}$ & 0.13 $\pm$ 3.6 $\times10^{-3}$ & 0.44 $\pm$ 1.8 $\times10^{-3}$ \\

\end{tabular}
\caption{\label{t:OdownHdown} The probabilities for finding water molecules with O-down (both hydrogens further away from the surface), O-up (both hydrogens closest to the surface), or other in the surface layers of water next to the metal surfaces. Here, $\sigma$ is the standard error of the mean.}

\end{table}
\end{center}

\begin{figure}[h!]
\centering
\includegraphics[width=\linewidth]{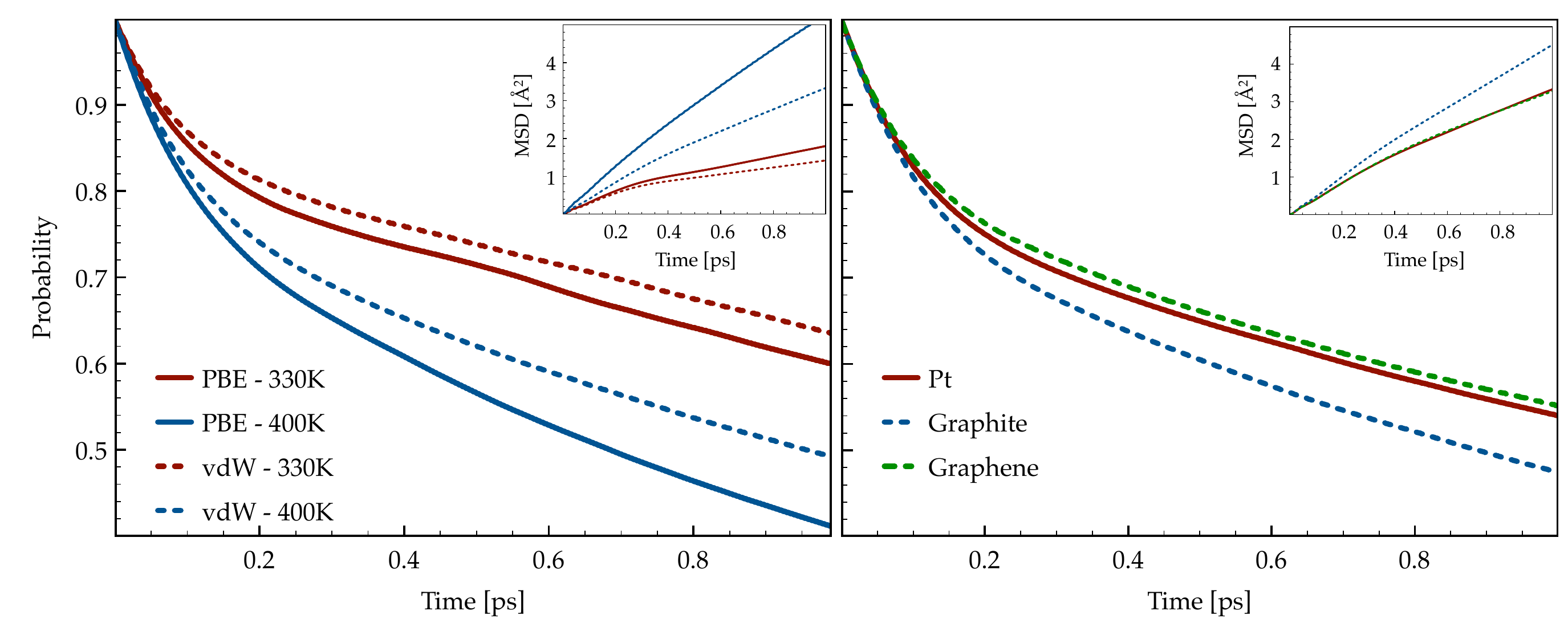}
\caption{\label{f:nrt_msd} Network reorganization probability functions ($\Pi(t)$) giving the likelihood of finding the same neighbouring water molecules as a function of time, and the mean squared displacements (subfigures) for selected simulations. In the left plot, we compare the liquid water-Pt interfaces with vdW interactions (labelled as vdW) and without (labelled as PBE) as well as temperature effects (330 and 400 K). In the right plot we compare the liquid water-Pt (labelled as Pt), liquid water-graphite (labelled as Graphite), and liquid water-graphene (labelled as Graphene) interfaces, all with a thermostat of 330 K using the optB86b-vdW exchange correlation functional.}
\end{figure}

Next, we examine the dynamics of the water molecules. In Figure \ref{f:nrt_msd}, we plot the mean squared displacement and a probability distribution we call the network reorganization probability function which we label as $\Pi(t)$. $\Pi(t)$ gives the likelihood of finding the same neighbouring water molecules as a function of time. This type of function has been used in previous literature to study pure hydrogen \cite{PhysRevLett.104.065702}, and gives similar distributions to water dipole rotational autocorrelation functions reported by Cicero \emph{et al.} \cite{cicero2008water}. For the mean squared displacements, there is a clear difference between the optB86b-vdW and PBE simulations. The use of the PBE functional causes the water molecules to be much more mobile, and therefore increases the likelihood of breaking bonds as water molecules diffuse away. For $\Pi(t)$, the temperature difference for both the optB86b-vdW and PBE simulations is identical (and expected). In Figure \ref{f:nrt_msd}, we can see that the likelihood of finding the same nearest neighbour decreases as you increase temperature. As the temperature is increased, more energy is introduced into the system which causes the kinetic energy of the water molecules to be greater than the energy associated with forming an intermolecular hydrogen bond. This explains why the curves are shifted.

\section{Comparing common electrodes}
In this section, we compare the liquid water-Pt interface with the liquid water-graphite and liquid water-graphene interfaces (using the optB86b-vdW exchange correlation functional). Graphene has shown extraordinary structural and electronic properties due to its massless Dirac fermions \cite{novoselov2005two}. As mentioned previously, an exceptional larger scale water-graphene theoretical study was done by Cicero {\it et al.} \cite{cicero2008water} using a PBE functional (without vdW interactions) to examine the overall structure and hydrogen-bonding network in closer detail. In another report, Li \emph{et al.} \cite{li2012influence} studied how the adsorption of a single water molecule effects the electronic structure of metal-supported graphene. Here, they used DFT with vdW interactions (using optB86b-vdW) and found that the $\pi$ and $\pi^*$ bands of graphene are not strongly perturbed by water adsorption in the case of a strong graphene-metal contact. Due to the stacked layers of graphite interacting solely through vdW forces, it can only be described correctly using non-local exchange correlation functionals. It is only the recent development of vdW exchange correlation functionals that allows for an accurate description of graphite in the DFT framework.

\begin{figure}[h]
    \centering
\includegraphics[width=0.5\linewidth]{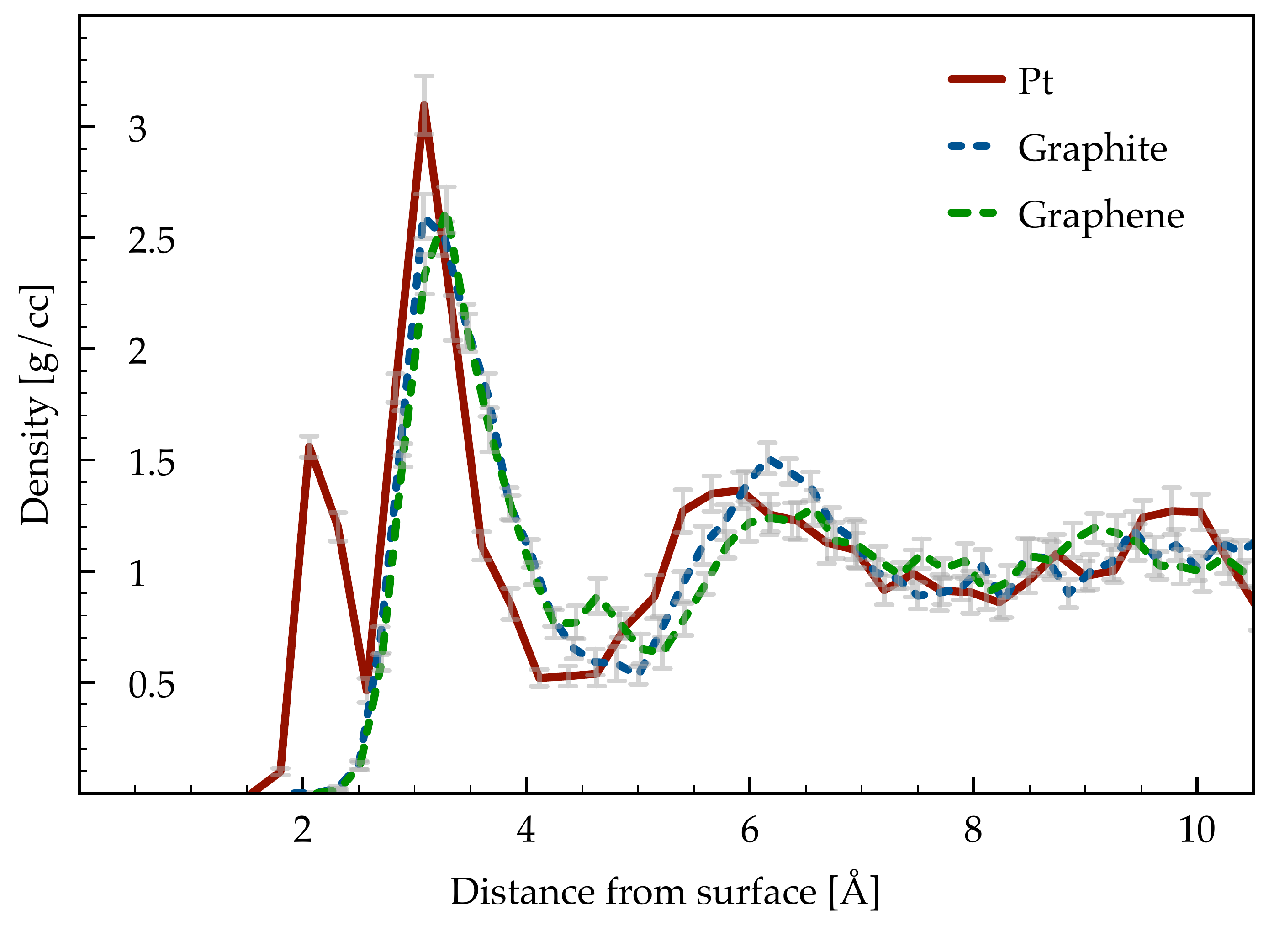}
\caption{\label{f:pt_graphite_comparison} Molecular density profiles for the liquid water-Pt (labelled at Pt), liquid water-graphite (labelled at Graphite), and liquid water-graphene (labelled as Graphene) interfaces. For all three interfaces, the simulations were run with a thermostat of 400 K, and the optB86b-vdW exchange-correlation functional was used.}
\label{f:PDOS}
\end{figure}

The first notable difference between the Pt and the graphite/graphene density profiles (Figure \ref{f:pt_graphite_comparison}) is the initial peak next to the Pt surface ($\approx 2~$\AA~ from the surface). This peak is absent for both the graphite and graphene interfaces. Given the adsorption geometries and band structures of single water molecules adsorbed to the surfaces of Ru(0001), Rh(111), and Pd(111) \cite{carrasco2009insight}, we suspect a chemisorption peak to be present for these surfaces. For Pt, water is well known to chemisorb to the surface, and the subsequent structural minimization calculation we performed for a single water molecule next to the Pt surface confirmed that a covalent bond forms on the atop site. For graphite, water does not chemisorb. When we structurally minimized the system with a single water molecule on the graphite surface, the water molecule is approximately 1 \AA~further away from the surface than the single water molecule sitting on a Pt slab, and the preferred adsorption site is the hollow site. This is consistent with the work done by Ambrosetti \emph{et al.} \cite{ambrosetti2011adsorption}. They found that the optimal bond length for a single water molecule on a graphite surface is greater than 3.2 \AA, with the water molecule sitting on the hollow site. When calculating the adsorption energy of water on graphite, Ambrosetti \emph{et al.} found very weak adsorption values ($63.5 \leq |E_{\text{ads}}| \leq 143.8$ meV depending on the exchange-correlation functional). In the case of Pt, binding energies on the Pt surface (-403 meV  \cite{carrasco2013role}) are almost 3 times the binding energy on graphite when using a vdW exchange correlation functional (-143.8 meV \cite{ambrosetti2011adsorption}). When examining the partial density of states (PDOS) of the $p$ orbitals on the O atom, and the $p_z$ band of the graphite surface atoms, it looks like a bond would be favourable given the overlap of energies of the $1b_1$ (highest occupied $p$ orbital) molecular orbital and the $p_z$ band (Figure \ref{f:PDOS}). The Pt $d$ band and the $1b_1$ orbital of water also overlap. The $1b_1$ orbital couples with the $d$ band causing the peak to broaden in the distribution. Although the energies of the $p_z$ orbitals and the $1b_1$ orbital coincide for the graphite slab, there is no coupling between the states (no broadening in the distribution for the $1b_1$ orbital). For the liquid water-graphite and liquid water-graphene density profiles, the largest peaks occur at about 3~\AA~away from the surface, consistent with the density profiles reported by Cicero \emph{et al.} for bulk water next to a graphene slab at 330 K \cite{cicero2008water}. The peaks are slightly shifted in comparison to the density profile of the liquid water-Pt interface, consistent with the adsorption geometries of the single water molecules next to the surfaces. Comparing the density profile of the liquid water-graphite interface with the liquid water-graphene interface, the initial peak in the density profile of the water-graphite interface is slightly larger than the initial peak of the water-graphene interface. This indicates that the extra layers in graphite increase the attraction of water molecules to the surface. This is consistent with Ambrosetti \emph{et al.} \cite{ambrosetti2011adsorption}, where they considered a single water molecule next to graphite and graphene and found the water molecule sits closer to the graphite surface.
 
\begin{figure}[h]
    \centering
\includegraphics[width=\linewidth]{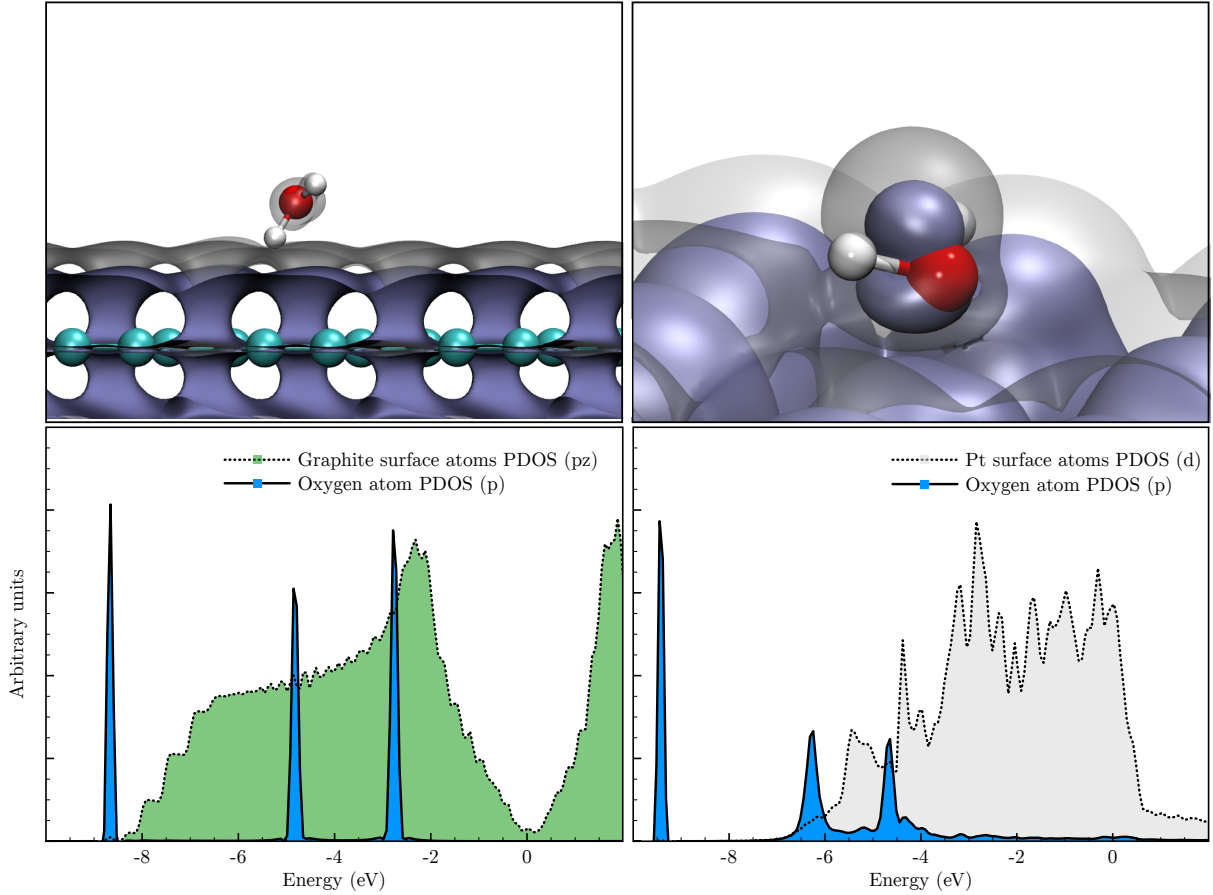}
\caption{\label{f:pt_graphite_PDOS} (Colour online) Visualizations of the HOMO orbital (top) for the graphite (left) and Pt (right) surface as well as the PDOS (bottom) for the graphite $p_z$ orbitals, Pt $d$ orbitals, and O $p$ orbitals. The opaque purple surface (1.1e-5) and the transparent black surface (8.2e-5) are at different isovalues.}
\end{figure}

When comparing the orientation of the water molecules (Figure \ref{f:pt_orient}), the water molecules next to the graphite surface show little preferred orientation (this was identical for the graphene interface). The maximum of the distribution gives an O-down orientation but the absence of peaks or valleys in the distribution suggests no preferred orientation. When Cicero \emph{et al.} \cite{cicero2008water} examined the orientation of bulk water next to graphene, they found that the preferred orientation was one hydrogen pointing down towards the surface, with the other hydrogen contributing to the local hydrogen bonding network in the surface layer. This is consistent with the geometry of the single water molecule sitting on the graphite surface, where we found this orientation energetically preferable. Looking to Table \ref{t:OdownHdown}, the probabilities of finding O-down, O-up, and mixed orientations of water molecules in the surface layer have the same trend when next to the Pt slab. The maximum of the orientation distribution most likely arises due to the water molecules closest to the surface, where 50\% of the time the oxygen faces the surface with both H atoms participating in the hydrogen bonding network.

Next, we compare the dynamics of the water molecules. Looking to Figure \ref{f:nrt_msd}, the mean squared displacements are similar, but the diffusion of water molecules is greatest for the graphite interface, and almost identical for the liquid water-Pt and liquid water-graphene interfaces. As in the case with the mean squared displacements, the probability functions, $\Pi(t)$, produces similar curves for all 3 interfaces. The hydrogen bonding network is most active for the liquid water-graphite interface and again almost identical for the liquid water-Pt  and liquid water-graphene interfaces. In this case, a more active hydrogen bonding network implies that intermolecular hydrogen bonds break and new bonds form (with other molecules) at a faster rate. In the work done by Cicero {\it et al.} \cite{cicero2008water}, they computed a water dipole rotational autocorrelation function which dropped off exponentially as a function of time. The faster the decay of the distribution indicates more mobility and reorientation of the water molecules. Interestingly, these dipole rotational autocorrelation functions are remarkably similar to $\Pi(t)$. This indicates a strong correlation between nearest neighbour dynamics and the rotational autocorrelation function of the dipole vector of a water molecule. Physically, this is intuitive as the orientation of a water molecule is highly dependent on the neighbouring water molecules. When the orientation of a water molecule drastically changes due to thermal fluctuations, it is possible that a new hydrogen bond will form with a different neighbouring water molecule. This would cause local restructuring in the hydrogen bonding network, in turn causing the dipole vectors of the local water molecules to reorient themselves.

\section{The effects of defects and dopants in graphitic surfaces}
Current experimental growth mechanisms of graphitic surfaces sometimes lead to concentrations of defects \cite{banhart2010structural}. These sites break symmetry in the lattice, leading to extra electrons or holes, which may lead to an increase in surface activity. In addition to increasing surface activity of inert materials via defects, many studies involving graphitic surfaces and dopant materials have been done to try to improve the catalytic performance of the electrode \cite{panchakarla2009synthesis, zheng2013two, reddy2010synthesis, panchakarla2010boron}. In particular, nitrogen and boron doped graphene has been shown to exhibit p-type and n-type semiconducting behaviour \cite{panchakarla2009synthesis}. To explore both of these scenerios, we have completed a preliminary study of liquid water-graphene and liquid water-graphite surfaces where single point defects, Stone-Wales defects, and dopant materials (B and N) have been introduced at the surface. At a temperature of 400 K, we found that for most of the simulations, the structure and dynamics of the liquid water subtly change when the defects/dopants are introduced. Interestingly, for the Stone-Wales defects simulations of the liquid water-graphene interface, we found a substantial decrease in the mobility of the water molecules as well as an almost random orientation of water molecules at the surface in comparison to all other pristine graphitic surfaces analyzed. As presented previously, the orientation of water molecules next to the pristine graphitic surfaces is mostly random, with a slight preference for having one oxygen facing towards the surface. When the Stone-Wales defect is introduced into the graphene surface, this preferred orientation disappears, leading to a mostly random orientation of water molecules at the surface. The absence of a preferred orientation indicates the lack of surface activity, or even repulsion at the surface. This repulsion pushes the surface water molecules towards the bulk, and therefore confines the liquid water leading to the decrease of the mobility of the water molecules. This is in agreement with \cite{lu2005stone}, where they found that Stone-Wales defects in single-wall carbon nanotubes are less reactive than the pristine counterpart. The Stone-Wales defect in the graphite surface did not produce the same effects as the graphene surface. The structure and dynamics of the water molecules is almost identical with pristine graphite.

\section{Conclusion}
We have carried out large-scale molecular dynamic simulations for bulk water next to Pt, and doped/defect graphitic surfaces in ambient and high temperature conditions. After analysis of the density profile, we find water next to a Pt slab forms a chemisorption layer before the largest density peak. This feature was absent for the liquid water-graphitic interfaces, and has not been discovered in other studies of other water-solid interfaces (including a liquid water-Au interface). Noticing the disappearance of this chemisorption peak at 1000 K for the liquid-water Pt interface, subsequent non-equilibrium calculations were run at temperatures in the range of 500-800 K. It was concluded that at 650 K, the peak begins to disappear. At 800 K, the peak has almost fully disappeared. When comparing the exchange correlation functional choice (i.e. with and without vdW interactions), there are differences. In the density profile, the highest peak can be 15-29\% larger when the vdW interactions are included (depending on the temperature). The orientation of water molecules next to the surface have subtle differences but show similar structure, and the dynamics of the water molecules indicate a further attraction to the surface and less mobility when the vdW interactions are included. 

When comparing the orientation of a single water molecule next to the Pt surface and the orientation of surface molecules in bulk water next to the Pt slab, there is a significant difference. For a single water molecule, the oxygen faces towards the atop site, with the geometric dipole of the molecule almost parallel to the surface. When analyzing the orientation of surface molecules in the bulk liquid, this is not always the case. The surface molecules interact with the metal atoms, as well as with other water molecules in their local environment. This, along with thermal fluctuations, cause the surface water molecules to constantly reorientate themselves in the competitive force field. For a single water molecule next to the graphite/graphene surface, there is little attraction to the surface. The optimal structure has a single hydrogen pointing down toward the surface and the distance between the water molecule and the surface is almost 1 \AA~greater than the distance between a single water molecule next to a Pt slab. The surface molecules in the bulk liquid next to graphite/graphene show little preference for orientation, confirming the weak interaction between water molecules and the graphite/graphene surface. Although the PDOS of the $p_z$ band for the surface C atoms overlaps with the $p$ orbitals of O, there is no coupling between them. The PDOS of the $d$ band for the surface atoms of Pt also overlaps with the $p$ orbitals of O, and the states do couple. The $p$ orbital of O, which coincides with the 1b${}_1$ orbital of water, broadens when next to the Pt surface. 

When introducing defects and dopants into the liquid water-graphitic surfaces, we find, for the most part, almost no changes in the structure and dynamics of the water molecules. The only notable difference came from Stone-Wales defects in the graphene surface, where we found a decrease in the mobility of water molecules. We propose this mobility decrease coincides with the decrease of surface activity, inevitably causing the surface water molecules to repel from the surface. Future work includes investigating other materials and optimizing the search for observables indicative of an efficient catalyst. When considering a new metal for water electrolysis, one should expect to observe chemisorption as well broadening in the peaks of the $p$ orbitals in the PDOS of liquid water as an indication of an efficient catalyst.

\section{Acknowledgments}
The authors acknowledge NSERC, SOSCIP, and Compute Canada for funding and computational resources.
\bibliography{refs}

\end{document}